# International Scientific Collaboration of China: Collaborating Countries, Institutions and Individuals

Xianwen Wang[1,2]*, Shenmeng Xu[1,2], Zhi Wang[1,2], Lian Peng[1,2], Chuanli Wang[1,2]

[1]WISE Lab, Faculty of Humanities and Social Sciences, Dalian University of Technology, Dalian 116085, China.

[2]School of Public Administration and Law, Dalian University of Technology, Dalian 116085, China.

* Corresponding author.
Email address: xianwenwang@dlut.edu.cn
Website: xianwenwang.com

## Abstract

Using bibliometric methods, we investigate China's international scientific collaboration from 3 levels of collaborating countries, institutions and individuals. We design a database in SQL Server, and make analysis of Chinese SCI papers based on the corresponding author field. We find that China's international scientific collaboration is focused on a handful of countries. Nearly 95% international co-authored papers are collaborated with only 20 countries, among which the USA account for more than 40% of all. Results also show that Chinese lineage in the international co-authorship is obvious, which means Chinese immigrant scientists are playing an important role in China's international scientific collaboration, especially in English-speaking countries.

## Introduction

China produces the world's second greatest number of papers yearly now, only after the USA. Accordingly, much attention has been paid to the bibliometrics analysis of Chinese publication records recently, among which Ronald N. Kostoff has made the most attempts and published a series of papers on Chinese science and technology performance by quantitative analysis and assessment on SCI papers of China. According to his study of total Chinese publication output in Science Citation Index/ Social Science Citation Index (SCI/SSCI), China's output of research articles has significantly expanded from 1996 to 2005, especially in cutting-edge technologies, such as nanotechnology and energetic materials. Compared to the USA, the bulk of China's articles focus on the physical and engineering sciences, while the USA articles focus on medical, social, and psychological sciences (Kostoff *et al.*, 2007a). The relative performance of science and technology in the USA and China was also compared in terms of quantity and quality in related studies (Kostoff, 2008; Kostoff, 2009).

As for the comparison of Chinese and Indian, Kostoff's study shows brief pictures of the Indian and Chinese S&T establishments (Kostoff *et al.*, 2007b), stating that China's rapid publication growth rate is continuing. His another paper goes deep into the technical organizations, the main technical thrusts, the characteristics of the main publication journals, the impact of collaboration on publication quality, the main technical areas of emphasis, and how citations compare by technical categoryind. (Kostoff, Bhattacharya & Pecht, 2007).

Other scientists, such as Loet Leydesdorff, Ping Zhou, have also done a lot of quantitative research on Chinese science and technology development (Leydesdorff, 2011; Leydesdorff & Wagner, 2009; Zhou & Leydesdorff, 2006). The citation rate of papers with Chinese corresponding authors shows an exponential growth (Zhou, 2008). More specifically, China has become a major player in critical technologies like nanotechnology (Zhou & Leydesdorff, 2006). The dynamics and the national characteristics of China's co-operation in a global context are also analyzed (Zhou & Glänzel, 2010).

Other research investigates Chinese performance in scientific publications at the macroscopic level. It can be indicated that many Chinese labs have made rapid progress according to qualitative assessment of fields of research and development. China will soon rival the others as a scientific superpower in many indicators (Shelton & Foland, 2009).

Chinese science and technology development in some specific scientific fields, such as nanoscience and nanotechnology, are also attracting increasing interest and attention. Bibliometric studies are conducted on the developing trends of nanotechnology, including both publications in nanotechnology field and nanotechnology patent applications. According to the research results, now China has become a nanoscience 'giant' (Kostoff, Barth & Lau, 2008; Guan & Ma, 2007; Liu *et al.*, 2009).

In addition, studies on different levels reveal a lot of information and significant results.

On country level, Jin & Rousseau (2005) observe the exponential growth of internationally co-authored papers of China. Tianwei He's results indicate that international collaboration publication output between China and the G7 countries has shown exponential growth aroused by the growth of science in China, and notably, USA is the most important collaborative country for China (He, 2009).

On institution level, Li Tang and Philip Shapira's research focus on the China–US scientific collaboration in nanotechnology. Through the collaboration analysis of institutions, they conclude that "The pattern of China's nanotechnology R&D collaboration with the US is asymmetrical, with a relatively small number of elite Chinese research organizations and universities working with a wide array of US universities" (Tang & Shapira, 2011).

On individual level, collaboration of individual scientists between China and USA plays an important role in China-US scientific collaboration, and the role of Chinese-American scientists is especially important (Wang *et al.*, 2012).

In spite of all mentioned above, most previous studies have certain defects and limitations. Firstly, due to the particularity of Hong Kong and Macao, they are not supposed to be included in the international cooperation data of China. Secondly, when calculating the number of Chinese papers, some research counts all the collaborators, not only first author or corresponding authors. As far as we are concerned, however, the counts dealing only with first author or corresponding authors can better reveal collaborations because generally speaking, they play a more leading role in scientific research cooperation. Thirdly, few previous studies go deep into the level of individual scientists. Analyses at the macroscopic level can't reveal enough details of science and technology structure, infrastructure and how they develop.

## Data and Methods

**Data sources**

Our data is collected from Web of Science. We search the data with the query *CU= China and PY=2010*. The citation database is restricted in Science Citation Index Expanded (SCI-EXPANDED), and only document types of article, letter, review, and editorial material are taken into account.

138,362 records are collected, including 133,218 articles, 1,066 letters, 2,763 reviews, 0 notes, and 1,315 editorial materials. Hong Kong and Macau are excluded, because as special administrative regions of China, Hong Kong and Macau is much more international than mainland China, which would cause large deviation to the result. And due to the "one country, two system" policy of China, this difference of international scientific collaboration between Mainland China and Hong Kong & Macau will last for a long time.

**Data processing**

Before the analysis begins, several problems need to be solved.

Firstly, we mainly focus on the papers with China appearing in the Reprint Addresses field. However, it is impossible to get these records directly. Consequently, we need to extract the records manually.

Secondly, as is mentioned above, we only focus on the Mainland China in this research. When we search Chinese SCI papers in Web of Science with the query *CU= China and PY=2010*, we find all the records with 'China' appearing in the Addresses field. So we need to extract the records of Mainland China. However, for those papers published by the mainland subsidiaries of Hong Kong institutions, they are kept in the database.

Thirdly, the format of some records is not standard. In some records, the Addresses field for authors is not complete. Sometimes, the Reprint Addresses field is also missing in some records. Accordingly, for these data, the format needs to be standardized. We check the nonstandard records and add all the missing pieces of information to the database one by one, according to the original full papers and their curriculum vitaes.

**Design of SCI paper database in SQL Server**

Considering there is no Reprint Addresses field in the Results Analysis Tool in Web of Science, to better analyze the data quantitatively, we design a database with SQL Server 2000 (Wang *et al.*, 2011a, Wang *et al.*, 2011b).

Every detailed field of WoS records is parsed, among which the most difficult part is the Addresses field. In most cases, there are more than one author and more than one address in one record, and sometimes some authors may have more than one address. In order to solve this problem, we need to link every author with his/her addresses in the database. Finally, total data of 138,362 records acquired from Web of Science are parsed and imported into our database.

## Results and Discussion

**Overall statistics**

5,784 records without Mainland Chinese authors are excluded, which means these papers are published only by authors from Hong Kong or Macau. Subsequently, there are 132,578 records left in the database, including 128,125 articles, 906 letters, 2,496 reviews, and 1,061 editorial materials.

Totally, our data includes 117,642 records with 'China' appearing in the Reprint Addresses field, accounting for 88.73% of all the 132,578 papers.

Generally speaking, the corresponding author plays a key role in initiating and organizing the research. In other words, in most cases, it is the corresponding author that directly forged the partnership. Looking deep into our data, the corresponding ratio for the article is 89%, and 88.19% for letter, 79.57% for review, 77.66% for editorial material. Obviously, the corresponding ratio for the review and editorial material records are lower compared to article and letter records.

Table 1 Numbers of Chinese SCI papers in 2010

| Document types | All papers | Corresponding papers | Corresponding ratio |
|---|---|---|---|
| Article | 128,125 | 114,033 | 89.00% |
| Letter | 906 | 799 | 88.19% |
| Review | 2,496 | 1,986 | 79.57% |
| Editorial Material | 1,061 | 824 | 77.66% |
| Total | 132,578 | 117,642 | 88.73% |

**International scientific collaboration analysis**

In this part, we focus on the international co-authorship of Chinese SCI papers based on the Reprint Addresses field in the records. The international co-authorship of Chinese SCI papers are classified into 2 groups: Chinese as corresponding authors, and foreigners as corresponding authors.

*Country Analysis*

*Top Collaborating countries with Chinese as corresponding authors* The total number of international co-authorship records with Chinese as reprint authors is 16,930. As Table 2 shows, the USA has collaborated with Chinese corresponding authors in 7,153 papers, accounting for 42.25% of all the 16,930 records. Japan has 1,681 collaborative papers, accounting for 9.93%. The United Kingdom (including England, Scotland, Wales, and North Ireland) has 1,476 papers co-authored with China, accounting for 8.72%.

The top 3 collaborating countries have participated 10,069 papers together (cumulated number), which account for 59.47% of all international co-authored papers with Chinese corresponding authors (cumulated ratio). And the top 10 foreign countries account for 88.62%, when the top 20 account for 94.88%. That means nearly 95% of international co-authored papers are collaborated with these 20 countries, and nearly half (42.25%) are with the United States.

Table 2 International co-authorship of Chinese corresponding authors

| Rank | Foreign Country | # Papers | % Papers | Cumulated # Papers | Cumulated % Papers |
|---|---|---|---|---|---|
| 1 | USA | 7153 | 42.25% | 7153 | 42.25% |
| 2 | Japan | 1681 | 9.93% | 8713 | 51.46% |
| 3 | UK | 1476 | 8.72% | 10069 | 59.47% |
| 4 | Australia | 1243 | 7.34% | 11168 | 65.97% |
| 5 | Canada | 1186 | 7.01% | 12188 | 71.99% |
| 6 | Germany | 1095 | 6.47% | 13100 | 77.38% |
| 7 | Singapore | 774 | 4.57% | 13763 | 81.29% |
| 8 | France | 693 | 4.09% | 14315 | 84.55% |
| 9 | South Korea | 513 | 3.03% | 14731 | 87.01% |
| 10 | Sweden | 317 | 1.87% | 15003 | 88.62% |
| 11 | Netherlands | 279 | 1.65% | 15215 | 89.87% |
| 12 | Italy | 188 | 1.11% | 15353 | 90.69% |
| 13 | Switzerland | 166 | 0.98% | 15461 | 91.32% |
| 14 | Russia | 164 | 0.97% | 15569 | 91.96% |
| 15 | Spain | 144 | 0.85% | 15672 | 92.57% |
| 16 | Belgium | 142 | 0.84% | 15763 | 93.11% |
| 17 | India | 115 | 0.68% | 15763 | 93.11% |
| 18 | Denmark | 106 | 0.63% | 15835 | 93.53% |
| 19 | New Zealand | 103 | 0.61% | 15918 | 94.02% |
| 20 | Norway | 99 | 0.58% | 16063 | 94.88% |

*Top collaborating countries with Chinese as participating authors* China has participated 13,309 papers where other countries appear in Reprint Addresses field. As Table 3 shows, the number of co-authored papers of the USA and China is 5,418, accounting for 40.70% of all the 13,309 records. Japan ranks second with 1,217 corresponding papers which have Chinese scientists as participators, and the proportion is 9.14%. Germany has 794 corresponding papers collaborated with Chinese authors, and the proportion here is 5.97%. The top 3 countries have 7,428 corresponding papers in total (cumulated number), and the cumulated proportion is 55.81%. What's more, the cumulated number of co-authored papers for the top 10 countries is 11,439, and the cumulated proportion is 85.95%. For the top 20 countries, the cumulated number is 12539, and the cumulated proportion is as high as 94.21%.

Table 3 International co-authorship of foreign corresponding authors

| Rank | Foreign Country | # Papers | % Papers | Cumulated # Papers | Cumulated % Papers |
|---|---|---|---|---|---|
| 1 | USA | 5417 | 40.70% | 5417 | 40.70% |
| 2 | Japan | 1217 | 9.14% | 6634 | 49.85% |
| 3 | Germany | 794 | 5.97% | 7428 | 55.81% |
| 4 | Canada | 792 | 5.95% | 8221 | 61.77% |
| 5 | UK | 777 | 5.84% | 8998 | 67.61% |
| 6 | Australia | 667 | 5.01% | 9665 | 72.62% |
| 7 | South Korea | 585 | 4.40% | 10250 | 77.02% |
| 8 | Singapore | 494 | 3.71% | 10744 | 80.73% |
| 9 | France | 448 | 3.37% | 11191 | 84.09% |

| 10 | Sweden | 248 | 1.86% | 11439 | 85.95% |
| 11 | Italy | 194 | 1.46% | 11633 | 87.41% |
| 12 | Netherlands | 179 | 1.34% | 11812 | 88.75% |
| 13 | Switzerland | 134 | 1.01% | 11946 | 89.76% |
| 14 | Spain | 114 | 0.86% | 12060 | 90.62% |
| 15 | Belgium | 107 | 0.80% | 12167 | 91.42% |
| 16 | Russia | 99 | 0.74% | 12266 | 92.16% |
| 17 | Denmark | 73 | 0.55% | 12339 | 92.71% |
| 18 | Austria | 70 | 0.53% | 12409 | 93.24% |
| 19 | Norway | 67 | 0.50% | 12476 | 93.74% |
| 20 | Finland | 63 | 0.47% | 12539 | 94.21% |

Fig. 1 compares the number of Chinese corresponding papers with the numbers of other top 10 countries together. For most countries, the numbers of collaborative papers with China as the corresponding countries (the left column) is greater than the numbers with other countries' corresponding papers (the right column). However, South Korea is an exception, which has more corresponding papers than China for the co-authored papers.

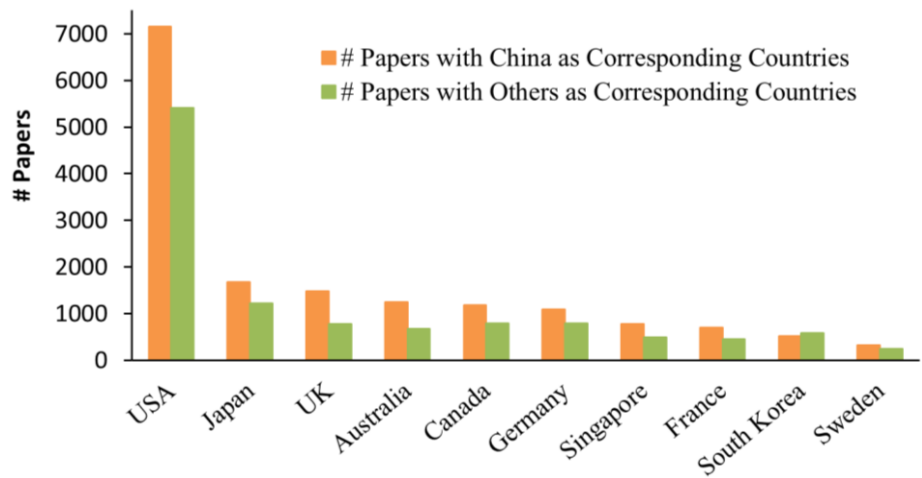

Fig. 1 Comparison of the number of papers with China or others as corresponding countries

*Institutional analysis*

For the total 16,930 international co-authored papers with Chinese as reprint authors, Zhejiang University has the most records of 615 papers, Peking University has the second most records of 571 papers, and Tsinghua University ranks third with 496 papers. Other Chinese top corresponding institutions are Shanghai Jiao Tong University, Fudan University, Nanjing University, etc.

Chinese Academy of Sciences (CAS) is the institution which has the most corresponding papers co-authored with other countries, and the number is 2383. Compared with normal Chinese universities that usually employs 3,000 to 5,000 researchers, CAS has over 50,000 researchers, which is 10 times as many as other universities in China. Consequently, in this study, CAS is divided into different parts according to its substitutions located in tens of cities in China.

For the 13,309 papers with authors in other countries appearing as reprint authors, Nanyang Technological University (Singapore) has the most corresponding papers of 224 co-authored with China. Other top foreign corresponding institutions are National University of Singapore (Singapore), Harvard University (the USA), Tohoku University (Japan), University of Michigan Ann Arbor (the USA), etc. Detailed information is listed in Table 4.

Table 4. Top corresponding institutions of China and other countries

| Rank | Chinese Corresponding Institutions | # Papers | Foreign Corresponding Institutions | # Papers |
|---|---|---|---|---|
| 1 | Zhejiang Univ | 615 | Nanyang Technol Univ | 224 |
| 2 | Peking Univ | 571 | Natl Univ Singapore | 191 |
| 3 | Tsinghua Univ | 496 | Harvard Univ | 92 |
| 4 | Shanghai Jiao Tong Univ | 451 | Tohoku Univ | 91 |
| 5 | Fudan Univ | 440 | Univ Michigan Ann Arbor | 85 |
| 6 | Nanjing Univ | 266 | Univ Calif Los Angeles | 85 |
| 7 | Harbin Inst Technol | 255 | Univ Tokyo | 82 |
| 8 | Xi An Jiao Tong Univ | 251 | Univ Calif Berkeley | 71 |
| 9 | Shandong Univ | 247 | Univ Florida | 67 |
| 10 | Univ Sci & Technol China | 246 | Univ Sydney | 62 |
| 11 | Huazhong Univ Sci & Technol | 225 | Univ Queensland | 62 |
| 12 | Sun Yat Sen Univ | 220 | Penn State Univ Univ Park | 61 |
| 13 | Dalian Univ Technol | 217 | Univ Wisconsin Madison | 60 |
| 14 | China Agr Univ | 208 | Ohio State Univ Columbus | 59 |
| 15 | Sichuan Univ | 208 | Univ Washington Seattle | 59 |
| 16 | Southeast Univ | 206 | Univ Alberta | 57 |
| 17 | Tongji Univ | 191 | Natl Inst Mat Sci | 56 |
| 18 | Jilin Univ | 184 | Univ So Calif | 56 |
| 19 | Beijing Normal Univ | 175 | Univ Texas MD Anderson Canc Ctr | 53 |
| 20 | Cent S Univ | 172 | Univ Pittsburgh | 53 |

As Fig. 2 shows, all the top 10 Chinese corresponding institutions have 3,838 co-authored papers, accounting for 22.67% of all the 16,930 international co-authored papers with Chinese as reprint authors. The top 20 Chinese corresponding institutions have co-authored 5,844 papers with international partners, accounting for 34.52%, when the number for the top 30 Chinese institutions is 7,251, with a proportion of 42.83%, and for the top 50, the number is 9,076 and 53.61%.

In contrast, the top 10 international institutions have only 1,050 corresponding papers participated by China, which account for only 7.69% of all the 13,309 papers with international corresponding authors and participated by China. And the number for the top 20 international corresponding institutions is 1,626, accounting for 11.91%. For the top 50 international institutions, the number is only 2,967, with a proportion of only about 21.72%.

In China, the top 50 Chinese institutions host over half of the international scientific collaboration of China, which means a relatively small number of elite Chinese institutions are working with a wide array of international institutions.

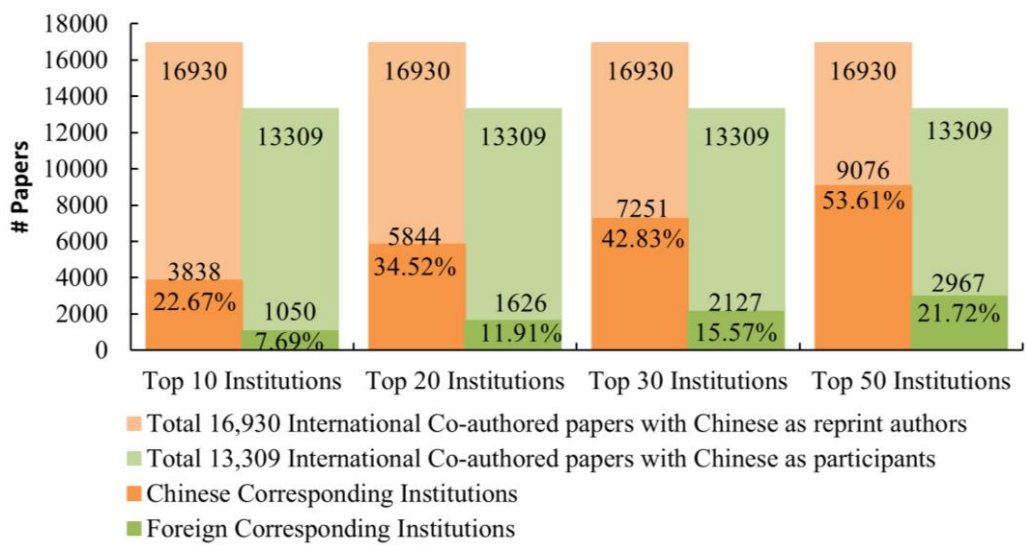

Fig. 2 Total records and percentage of top corresponding institutions

*Individual analysis*

To investigate the international co-authorship of Chinese papers in a microscopic view, we check the individual authors of the co-authored papers. According to the last names, we identify the authors with Chinese lineage. Usually, Chinese immigrants don't change their names when they move to other countries. Most Chinese last names are very identifiable, such as Zhang, Jiang, Guo, Xiong, etc. And Chinese phonetic (Pinyin) are unique. Usually, Chinese immigrants' names are written in Chinese Pinyin. In a few cases, some Chinese immigrants added English first names before/after their full names or replace their first names with English names, for example, Wang, Lei Phillip and Fan, Janet. However, combined with their last names and first names, their Chinese lineage are still very identifiable.

For some names that can't be identified from the spelling, it is necessary to be confirmed by searching and checking their personal information.

The Chinese lineage mentioned here refers to first-generation immigrants mostly. In a few cases, second-generation immigrants are also included. As a result, co-authored records in our study are checked one by one to identify the authors with Chinese lineage.

Fig. 3 shows the proportions of papers by authors with and without Chinese lineage. Here we analyze the Top 9 countries with the most papers in Table 3, which are the USA, Japan, Germany, Canada, UK, Australia, South Korea, Singapore and France. The size (area) of the pie chart illustrates the total co-authored number of papers. The grey sector indicates the proportion of records reprinted by authors with Chinese lineage, and the spotted sector represents the proportion of records reprinted by authors without Chinese lineage.

For the 5,417 papers with the USA as the corresponding country, we find that 3,497 records are reprinted by Chinese American authors, accounting for about 65% of all the 5,417 papers. For the 1,217 records with Japan as the corresponding country, about 32% (384 papers) are reprinted by Chinese Japanese authors. For Germany, the number for Chinese German is 220 and account for about 28% of all 794 papers. For Canada, Chinese Canadian reprinted 485 papers, which account for 61% of all. Fig. 3 shows the detailed information of these 9 countries.

Among these countries, for the USA, Canada, Australia and Singapore, all the proportions of papers reprinted by Chinese lineage authors are greater than 60%. The proportion for UK is about 48%. And for the other 3 countries, which are Japan, Germany and France, the proportions are about 30%.

Obviously, in most countries, authors with Chinese lineage play a very important role in Chinese international scientific collaboration, especially in the collaboration between China and English speaking countries. The high proportion of 75% in China-Singapore scientific collaboration and 65% in China-US collaboration have conformed this. For non-English speaking countries, such as Japan, Germany and France, the proportion of about 30% is also not low. However, South Korea is a special case. Although it is geographically close to China, the proportion is only 9%.

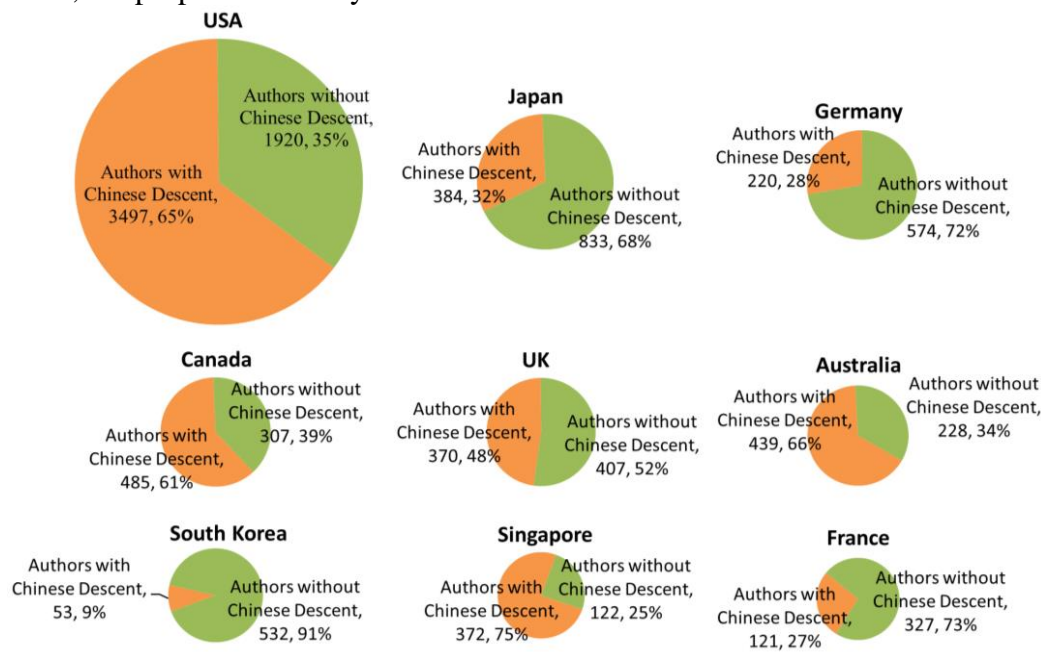

Fig. 3 The proportion of authors with Chinese Descent

## Conclusions

China's international scientific collaboration is focused on a handful of countries, among which the USA accounts for more than 40% of all. Other important countries are Japan, UK, Australia, Canada, Germany, Singapore, etc. The top 10 foreign countries account for more than 88% of all the collaborations, and the top 20 account for more than 94%, which means that nearly 95% international co-authored papers are collaborated with only 20 countries.

In addition, a relatively small number of elite Chinese institutions, including Zhejiang University, Peking University, Tsinghua University, *etc.*, host over half of the international scientific collaboration of China, and work with a wide array of international institutions.

An important new finding is that Chinese lineage in the international co-authorship is obvious, which means Chinese immigrants scientists are playing an important role in China's international scientific collaboration. In the collaborations with English-speaking countries such as the USA, Canada, Australia, Singapore and the United Kingdom, more than 45% of the foreign corresponding authors have Chinese lineage. Meanwhile, for the non-English speaking countries, such as Japan, Germany, France, the ratio is about 30%.

# Acknowledgement

The work was supported by the project of "Social Science Foundation of China"(10CZX011), the project of "Research Fund for the Doctoral Program of Higher Education of China"(20090041110001), and the project of "Fundamental Research Funds for the Central Universities" (DUT12RW309).